\documentclass{PoS}

\usepackage{graphicx,amsmath,amssymb,fontenc,times,mathptmx}

\title{Looking at the gluon moment of the nucleon with dynamical twisted mass fermions}

\ShortTitle{Looking at the gluon moment of the nucleon with dynamical twisted mass fermions}

\author{Constantia Alexandrou$^{ab}$, Vincent Drach$^c$, Kyriakos Hadjiyiannakou$^a$, Karl Jansen$^{ac}$, Bartosz Kostrzewa$^{d}$, \speaker{Christian Wiese}~$^c$\\
	\\
	$^a$Department of Physics, University of Cyprus, P.O. Box 20537, 1678 Nicosia, Cyprus\\
        \\
	$^b$Computation-based Science and Technology Research Center, Cyprus Institute, 20 Kavafi Str., Nicosia 2121, Cyprus\\
        \\
	$^c$NIC, DESY Zeuthen, Platanenallee 6, D-15738 Zeuthen, Germany\\
        \\
	$^d$Humboldt-Universit{\"a}t zu Berlin, Institut f{\"u}r Physik, Newtonstra{\ss}e 15, D-12489 Berlin, Germany\\
	\\
        E-mail: \email{christian.wiese@desy.de}}

\abstract{To understand the structure of hadrons it is important to 
know the PDF of their constituents, the quarks and gluons. 
In our work we aim to compute the 
first moment of the gluon PDF $\langle x \rangle_g$ for the nucleon. 
We follow two possible approaches in order to extract the gluon moment: 
the Feynman-Hellmann theorem and a direct method with smearing 
of the gluon operator. We present preliminary results 
computed on $24^3 \times 48$ lattices for the case where the Feynman-Hellman theorem
is used and $32^3 \times 64$ lattices for the direct method, employing 
$N_f=2+1+1$ maximally twisted mass fermions.}

\FullConference{31st International Symposium on Lattice Field Theory LATTICE 2013\\
		 July 29 - August 3, 2013\\
		 Mainz, Germany}

\begin{document}

\section{Introduction}
\vspace{-3mm}
From deep inelastic scattering (DIS) experiments it is known, see e.g. \cite{Devenish}, 
that the nucleon is not a 
fundamental particle but consists of so-called partons as its constituents. 
The contribution of these partons to the nucleon momentum is described 
by a parton distribution function (PDF) $f_p(x)$, which is the 
probability to find a parton $p$ with a momentum fraction $x$. 
The first moment of the PDF $ \langle x \rangle_p = \int x f_p(x) dx$ 
is the fraction of the total nucleon momentum carried by the parton. 
This implies then the energy-momentum sum rule $\sum_p \langle x \rangle_p =1$. 
The partons were eventually identified as the quarks and the gluons as
the fundamental building block of hadrons. Thus, the energy-momentum sum rule 
of partons translates directly to a sum rule involving all quarks and 
the gluons,
\begin{align}
  \sum_q \langle x \rangle_q + \langle x \rangle_g = 1\; .
\label{eq:sumrule}
\end{align}

Further experimental input suggests that  
the contribution coming only from 
up- and down quarks does not add up to one \cite{Blumlein:2006be}. Since it is expected that 
the heavier quarks will not significantly contribute to the average nucleon 
momentum \cite{Martin:2009iq}, this implies that the 
gluons carry a large amount of the nucleon momentum, such that the sum rule 
of eq.~(\ref{eq:sumrule}) is satisfied. 

Therefore, the computation of the gluon moment is necessary to fully 
understand the structure of the nucleon. However, at the moment, 
despite the fact that there are many results for the quark  
structure of the nucleon, see e.g. refs.~\cite{Alexandrou:2013joa, Alexandrou:2012gz}, 
there are just a few results for $\langle x \rangle_g$ which are,
moreover, only obtained from quenched computations \cite{Horsley:2012pz, Liu:2012nz}. 
The work presented here aims at giving a first result from a computation on gauge 
configurations generated with light, strange and charm sea quarks.

We can access the gluon moment of a hadron via the matrix elements of the gluon operator:
\begin{align}
O_{\mu\nu} = -\mbox{tr}_c G_{\mu\rho}G_{\nu\rho}\; .
\label{eq:operator}
\end{align}
 The matrix elements of this operator can be computed with a ratio 
of a three-point and a two-point function, where the sink time $t$ 
and the operator time $\tau$ are chosen properly.
\begin{align}
  \frac{\langle h(p,t)\mathcal O(\tau)h(p,0)\rangle}{\langle h(p,t)h(p,0) \rangle} \stackrel{0\ll \tau\ll t}= (\mathcal O)_{h(p)h(p)} 
\label{eq:me} 
\end{align}
where $h(p,t)$ denotes a hadron with momentum $p$ at sink time $t$. 
The general matrix element of eq.~(\ref{eq:me}) 
can be decomposed into several terms proportional to appropriate form factors, 
see e.g. the discussion in ref.~\cite{Alexandrou:2013joa}. 
The relevant form factor for our purpose is $A_{20}$, which can be 
related to the gluon moment. 
In order to proceed, we need to consider certain representations of the operator
in eq.~(\ref{eq:operator}). Here we choose two of them 
\begin{align}
  \mathcal A_{i}= \mathcal O_{i4},\;\;\mathcal B= \mathcal O_{44} - \frac{1}{3}\mathcal O_{jj}. 
\end{align}
The matrix elements of these operatoprs can be written in terms of the gluon moment as
\begin{align}
  \label{EQN_intro_2} 
  (\mathcal A_i)_{N(p)N(p)} = -i p_i\langle x \rangle_g,\;\;(\mathcal B)_{N(p)N(p)} = (m_N + \frac{2}{3E_N} {\vec{p}}\:^2)\langle x \rangle_g\; .
\end{align}
Both operators have certain drawbacks. 
The operator $\mathcal A_i$ can only be taken when a non-zero momentum 
is injected. It is known that the computation of momentum dependent operator matrix elements 
result in a larger noise-to-signal ratio than a momentum-zero computation, which is possible for
operator $\mathcal B$.

In case of the operator $\mathcal B$ there is a subtraction of 
two terms which are similar in magnitude. 
This can be understood from the lattice version of the operator, expressed in terms of plaquettes:
\begin{align}
  \label{EQN_intro_1}
  \mathcal B(t)= \frac{4}{9}\frac{\beta}{a}\sum_x\left(\sum_i\mbox{tr}_c[U_{i4}(x,t)]-\sum_{i<j}\mbox{tr}_c[U_{ij}(x,t)]\right)\; .
\end{align}
Here, one sees that the spatial and the temporal part of the plaquette, which are very similar 
in size, need to be subtracted, leading potentially again to a bad signal-to-noise behavior
of the corresponding matrix element. 
The choice we made for the following discussion is nonetheless  
the operator $\mathcal B$ since it is directly accessible to us. 

\section{Feynman-Hellman theorem}
\vspace{-3mm}
One approach to extract the matrix elements of the gluon operator 
that was applied in \cite{Horsley:2012pz} uses the Euclidean form of 
the Feynman-Hellman theorem. If one introduces some operator $\lambda \mathcal O$ 
into the action of the system, the operator's matrix elements 
can be derived from the derivative of the energy of the state with respect to $\lambda$.
\begin{align}
  \label{EQN_Feynman_1}
  \frac{\partial E_N(\lambda)}{\partial\lambda} = ( :\frac{\partial \hat{S}(\lambda)}{\partial \lambda}:)_{N(p)N(p),\lambda}\; .
\end{align}
Here $:...:$ means that the vacuum expectation value of the operator 
has to be subtracted. For the purpose of calculating the three-point function 
for the gluon operator we modify the Wilson gauge action as
\begin{align}
  S(\lambda) = \frac{1}{3}\beta(1+\lambda)\sum_i\mbox{tr}_c[1-U_{i4}]+\frac{1}{3}\beta(1-\lambda)\sum_{i<j}\mbox{tr}_c[1-U_{ij}]\; .
\end{align}
Note that $\lambda=0$ corresponds to the standard Wilson plaquette 
gauge action. When applying eq.~(\ref{EQN_intro_2}), 
(\ref{EQN_intro_1}) and (\ref{EQN_Feynman_1}) one can relate the 
derivative of the nucleon energy to $\langle x \rangle_g$. 
\begin{align}
  \frac{\partial E_N}{\partial\lambda} {\big |_{\lambda = 0}} = - \frac{3}{2}\left(m_N+\frac{2}{3E_n}\vec{p}\:^2\right)\langle x \rangle_g\; .
\end{align}
There is no subtraction of the vacuum expectation value here, 
because utilizing lattice rotational symmetry it can be 
shown that the expectation value of the operator in eq.~(\ref{EQN_intro_1}) is zero. 
When computing the nucleon mass at zero momentum, the relation can be simplified as:
\begin{align}
  \langle x \rangle_g = \frac{2}{3m_N}\frac{\partial m_N}{\partial \lambda}{\big |_{\lambda=0}}\; .
\end{align}
In order to compute the nucleon mass for different, non-zero $\lambda$ values, 
new gauge ensembles had to be generated. In addition, due to the change 
of the gauge action, the hopping parameter $\kappa$  had to be re-tuned 
to its critical value for each ensemble, in order to regain the 
$\mathcal{O}(a)$ improvement.

We have performed preliminary tests on small lattices with heavy quark masses 
to keep the computational effort affordable. The simulations were carried out 
with $24^3 \times 48$ lattices and $N_f=2+1+1$ flavors of 
maximally twisted mass fermions. We employed $\beta=1.95$ which corresponds to a 
lattice spacing of $a\approx 0.078$~fm and a twisted mass parameter $\mu=0.085$ 
which leads to a pion mass of $m_{PS} \approx 490$~MeV. As gauge action we 
used the Iwasaki action, however the Feynman-Hellman theorem was only applied to the 
Wilson part, i.e. the pure plaquette part, of the action.

\begin{figure}[htb]
  \centering 
  \includegraphics[scale=1]{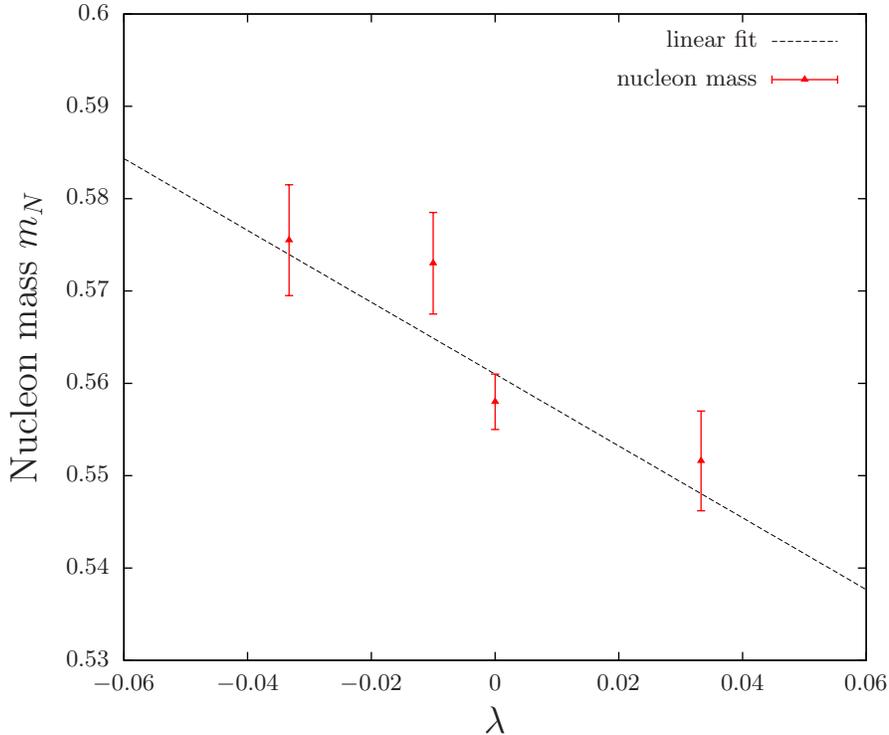}
  \caption{\label{FIG_FH_1}Dependence of the nucleon mass on the change of the 
gauge action (different $\lambda$ values). The slope of the fit can be related to the gluon moment.}
\end{figure}

Our results for three different $\lambda$ values on $\sim 200$ gauge 
configurations and the nucleon mass at $\lambda=0$ can be seen in Fig.~\ref{FIG_FH_1}.

We performed a 
linear fit in $\lambda$ to the data of the nucleon mass. The fact that the data 
shows a $\lambda$ dependence  suggests that we can obtain a non-zero signal for 
the gluon moment. However, the error of the slope is rather large (about 30\%). 
The systematic error is probably even larger, because it is not known in 
which $\lambda$ region a linear fit is really justified. To study this 
systematic effect one would need to compute the nucleon mass with a smaller 
error for more $\lambda$ points than used here.

\section{Direct method}
\vspace{-3mm}
An alternative, more straightforward method of computing the 
matrix element of eq.~({\ref{eq:me}) is a direct approach, where, 
through performing the relevant Wick contractions,  
the three-point function can be expressed by a suitable combination of 
propagators and gauge links.
For the gluon three-point function this is actually a trivial task, 
because there are no quark fields in the gluon operator. Subsequently, 
there are no possible contractions between the gluon operator 
and the interpolating fields of the nucleon. The three-point 
function can, in fact, be written as a product of nucleon two-point functions 
and the gluon operator. For the zero momentum computation we get 
\begin{align}
  \frac{\left\langle [N(t)N(0)]_{p=0} \mathcal B(\tau)\right\rangle}{\langle N(t)N(0)_{p=0} \rangle} \stackrel{0\ll \tau \ll t}=m_N \langle x \rangle_g\; .
\end{align}
The advantage of this method is that we can reuse existing two-point 
functions and only have to compute the gluon operator on the very same 
configurations which requires little computational effort.

The following results were computed on a $32^3 \times 64$ lattice 
with $N_f=2+1+1$ flavors of maximally twisted mass fermions. 
We set $\beta=1.95$, which corresponds to a lattice spacing of $a\approx 0.078$\;fm 
and the twisted mass parameter $\mu = 0.055$, which is a pion mass of 
$m_{PS} \approx 393$\;MeV. For the two-point function we used 16 different 
source positions on each gauge configuration which corresponds to 32 measurements, 
because we considered proton and neutron fields. The first results for a 
local gluon operator can be seen in the left panel of Fig.~\ref{FIG_DIR_1}. 

\begin{figure}[htb]
  \centering 
  \includegraphics[scale=0.7]{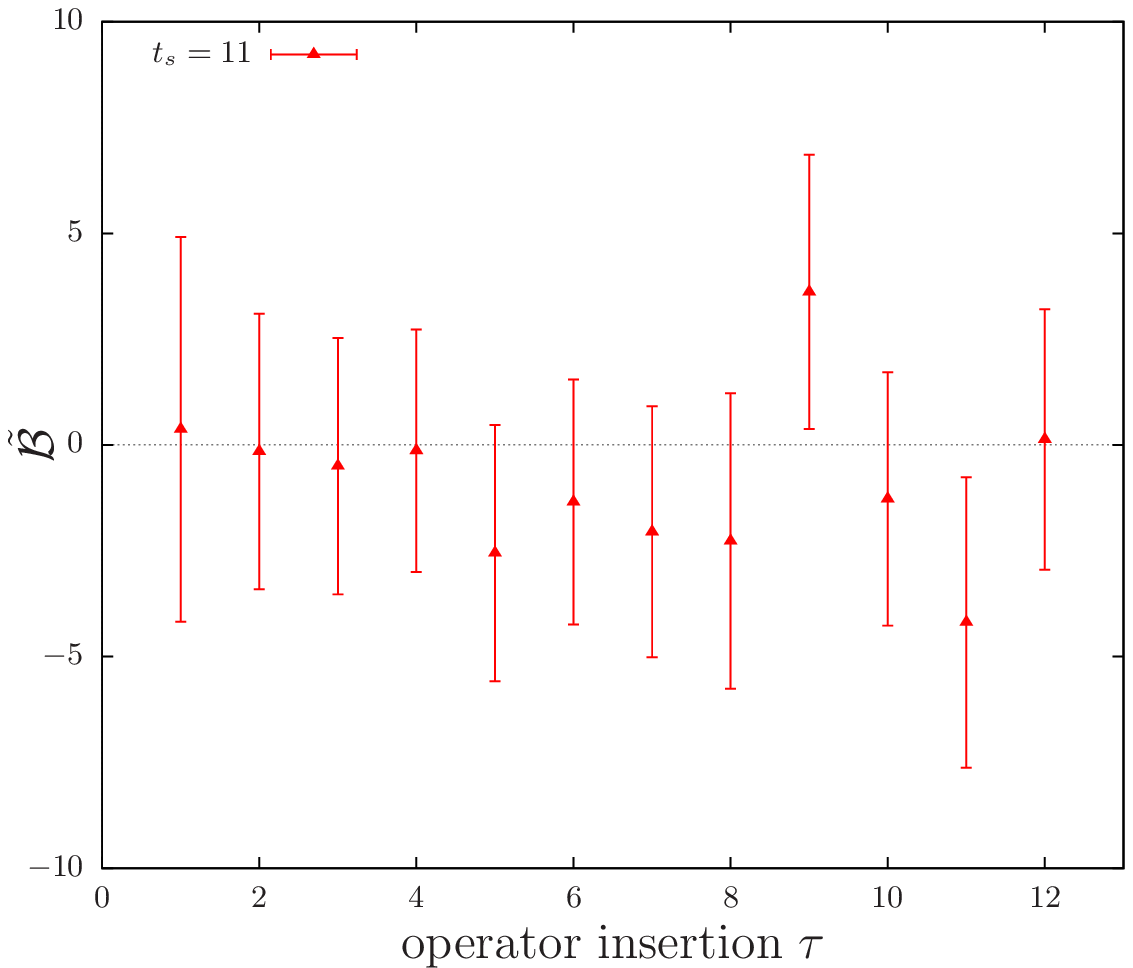}\;\;\includegraphics[scale=0.7]{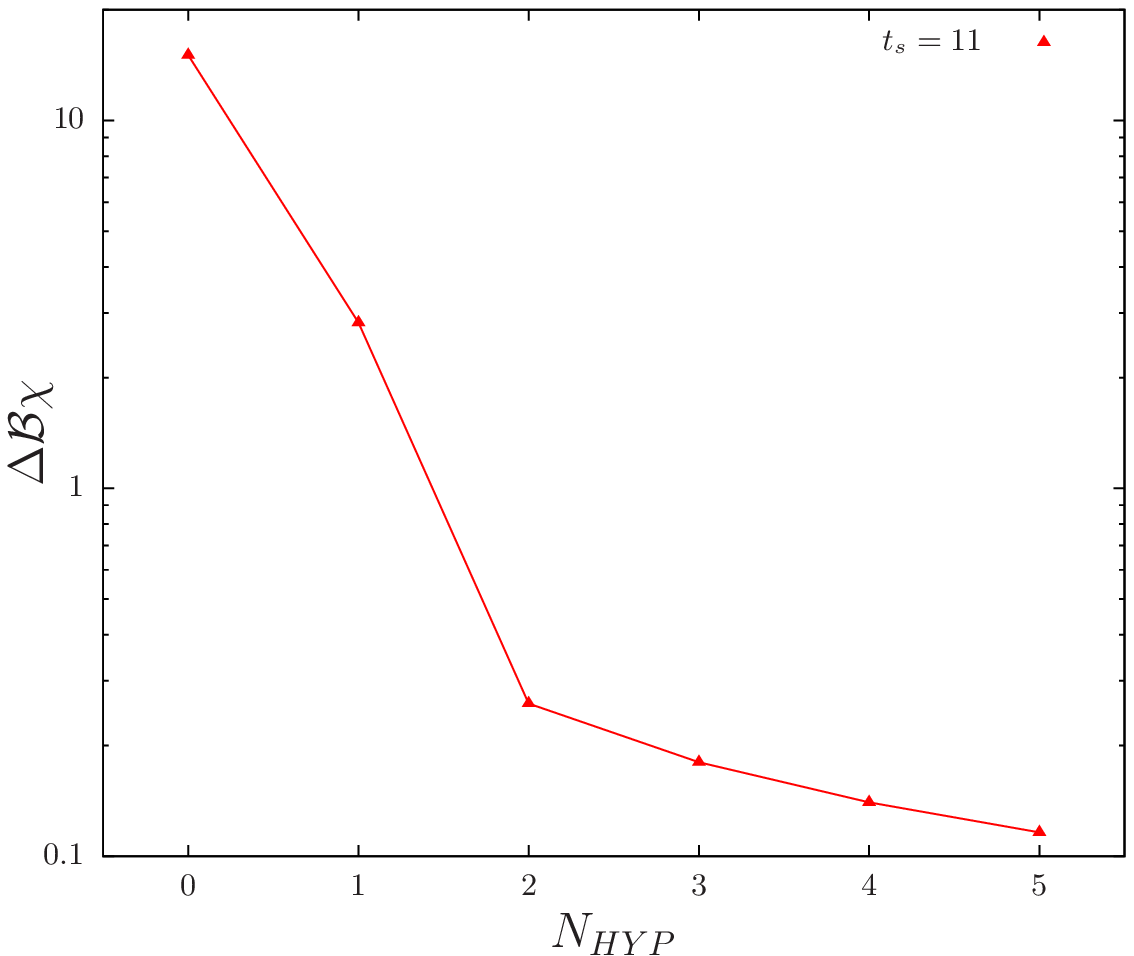}
  \caption{\label{FIG_DIR_1}{\bf left:} Nucleon matrix element for a local gluon operator for a source-sink separation of 11 and different operator insertion times $\tau$. {\bf right:} Relative error of the nucleon matrix element for different HYP-smearing steps of the gluon operator.}
\end{figure}

Obviously, it is not possible to extract a signal, due to a large noise-to-signal ratio. 
A possible solution for this problem can be found in \cite{Meyer:2007tm}, 
where it is suggested to use HYP smearing \cite{Hasenfratz:2001hp} for the links in the gluon operator.
We applied several steps of HYP smearing with parameters from \cite{Hasenfratz:2001hp} 
and present the relative error (noise-to-signal ratio) for the observable 
in the right panel of Fig.~\ref{FIG_DIR_1}.

We found a significant reduction of the noise-to-signal ratio with increasing number 
of HYP smearing steps. Thus, we subsequently applied five steps of HYP smearing.  

\begin{figure}[htb]
  \centering 
  \includegraphics[scale=0.7]{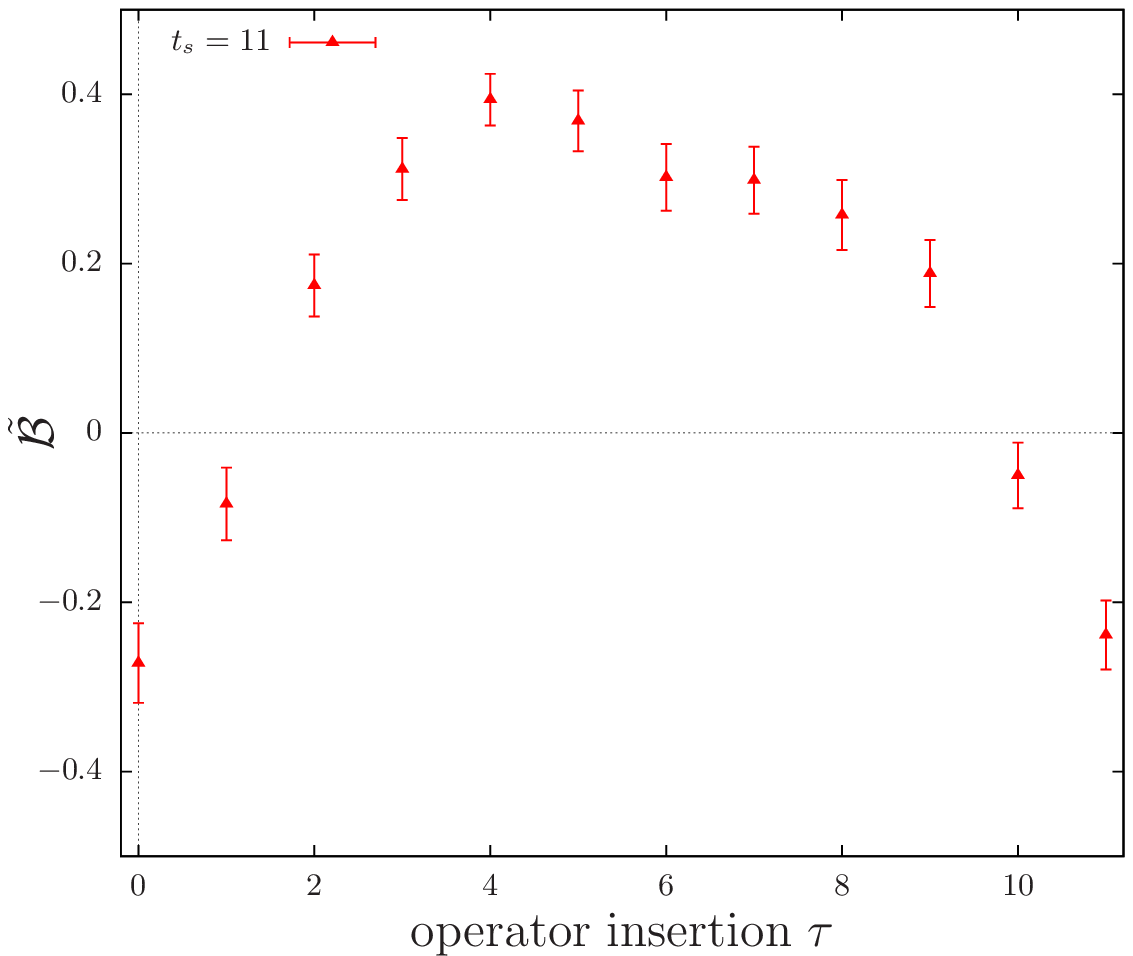}\;\;\includegraphics[scale=0.7]{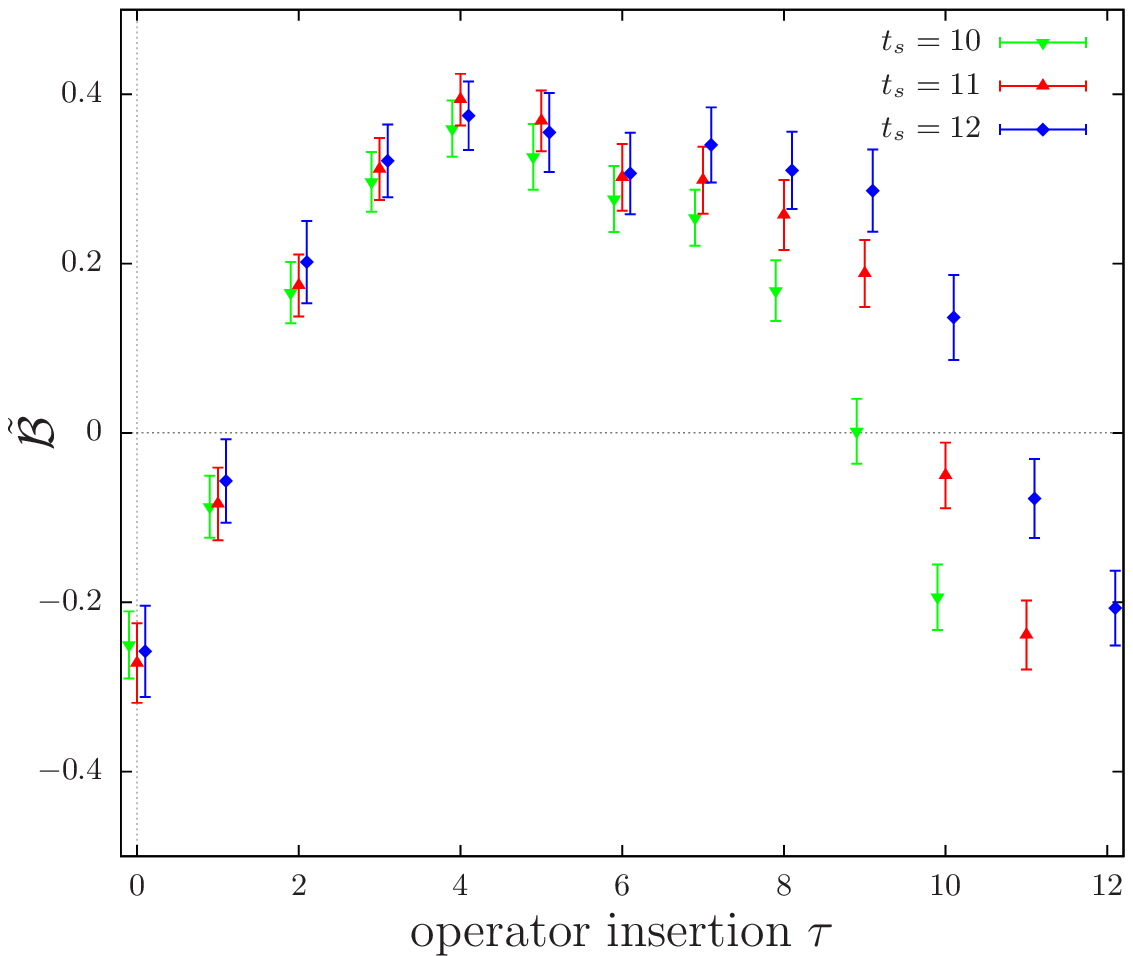}
  \caption{\label{FIG_DIR_2}{\bf left:} Nucleon matrix element for a HYP-smeared gluon operator for a source-sink separation of 11 and different operator insertion times $\tau$. {\bf right:} The same matrix element for three different source-sink separations. $\mathcal B=\frac{4}{9}\beta\chi\tilde{\mathcal B}$}
\end{figure}

On the left panel of Fig.~\ref{FIG_DIR_2} one can see the signal we got from a single source-sink 
separation, where $\mathcal B=\frac{4}{9}\beta\chi\tilde{\mathcal B}$ 
and $\chi$ is a normalization factor caused by using HYP smearing. We clearly 
got a non-zero value with a reasonable error of about 10\%. However, this 
signal could still be contaminated by excited state effects. This can be 
checked by computing the matrix element for different source-sink separations. 
On the right panel one can see that there are no strong excited states effects, 
because the plateau value seems to be stable for different sink time positions.

\section{Conclusion and outlook}
\vspace{-3mm}
We presented two methods which potentially can be used to extract $\langle x \rangle_g$ 
on the lattice: 
The first method makes use of the Feynman-Hellman theorem and has the advantage of yielding
a statistically significant signal for rather moderate statistics. However, the calculation 
needs dedicated simulations with different values of $\lambda$ to establish unambiguously the 
linear dependence of the results on $\lambda$. Furthermore, each 
simulation has to be tuned to a critical value of $\kappa$, in order to ensure automatic $\mathcal O(a)$ improvement. 
Therefore, overall, the computational cost associated with this method is large.

The second method directly computes the three-point function from which $\langle x \rangle_g$ can be extracted.
In order to obtain a non-zero signal, one has to apply smearing on the gauge links entering the operator.
Although one needs large statistics, one can use nucleon two-point functions computed for other observables
and therefore the overall cost is small.

Our study therefore suggests that the direct method may be the method of choice to calculate this particular observable.
Still, the Feynman-Hellman theorem could be used as a cross-check on ensembles where
it is feasible to apply.

Another issue regarding the computation of the physical value of $ \langle x \rangle_g$
is that the lattice matrix element needs to be renormalized.
Since the gluon operator is a singlet operator it will mix with the quark momentum fraction 
$ \langle x \rangle_q$. The relation between the renormalized and the bare 
values of both quantities is given by a $2\times2$ mixing matrix.
\begin{align}
  \binom{\langle x \rangle_g^{\overline{MS}}}{\sum_q\langle x \rangle_q^{\overline{MS}}} = Z_{2\times2}\binom{\langle x \rangle_g^{bare}}{\sum_q\langle x \rangle_q^{bare}} .
\end{align}
For $ \langle x \rangle_g$ the relevant matrix elements are called $Z_{gg}$ and $Z_{qq}$ and the relation is
\begin{align}
  \langle x \rangle_g ^{\overline{MS}} = Z_{bare\:gg}^{\overline{MS}}\langle x \rangle_g ^{bare} + [1-Z_{bare\:qq}^{\overline{MS}}]\sum_q \langle x \rangle_q ^{bare} .
\end{align}
As a first step we will compute these factors perturbatively. 
This will provide us with a first estimate of the factors and we will get 
insight in the renormalization process of this quantity. 
If we know the renormalization conditions, a non-perturbative renormalization can follow. 
Since the smearing of the operator should be included into the renormalization process, we will also try
to use other smearing techniques for the lattice computation, i.e. HEX or stout smearing, which can be easier employed 
in a perturbative computation.
Once the renormalization is complete we will be able to give the first physical 
result for $ \langle x \rangle_g$ with fully active sea quarks.

The next step should be the computation of $ \langle x \rangle_g$ 
at physical pion mass using the recently generated ensembles with 
$N_f=2$ twisted-mass-clover fermions \cite{ETMC}. 
For heavier quark masses the continuum limit for this quantity can be studied.

Furthermore, the result can be used for the determination of the gluon contribution to the nucleon spin, 
which at the moment is not known from the lattice. 
Moreover, it could also be possible to compute the gluon moment of other hadrons, e.g. the pion (cf.~\cite{Meyer:2007tm}).

\section*{Acknowledgments}
\vspace{-3mm}
We thank Carsten Urbach for discussion and help with the tmLQCD code \cite{Jansen:2009xp}, 
which has been used for the computations. 
Latest developlments for this code were also presented at this conference \cite{tmLQCD1,tmLQCD2}. 
This work has been supported in part by the DFG Sonderforschungsbereich/Transregio
SFB/TR9-03. B.K. is supported by the National Research Fund, Luxembourg.

\end{document}